\documentclass[9pt]{article}
\usepackage{spconfa4,amsmath,graphicx}
\usepackage{array}
\usepackage{epstopdf}
\usepackage{amssymb}
\usepackage{cite}
\usepackage{subfig}
\usepackage{empheq}
\usepackage{siunitx}
\usepackage[subtle,title=normal]{savetrees}

\newenvironment{nospaceflalign}
 {\setlength{\abovedisplayskip}{6pt plus 1pt minus 2pt}
 \setlength{\belowdisplayskip}{6pt plus 1pt minus 2pt}%
  \csname flalign\endcsname}
 {\csname endflalign\endcsname\ignorespacesafterend}
\newcommand{\be}{\begin{equation*}}
\newcommand{\ee}{\end{equation*}}
\newcommand{\bea}{\begin{eqnarray*}}
\newcommand{\eea}{\end{eqnarray*}}
\newcommand{\ba}{\begin{array}}
\newcommand{\ea}{\end{array}}
\newcommand{\lba}{\begin{bmatrix}}
\newcommand{\ear}{\end{bmatrix}}

\title{Optimal Binaural LCMV Beamforming in Complex Acoustic Scenarios:\\Theoretical and Practical Insights}
\name{Nico G\"o\ss ling$^1$, Daniel Marquardt$^{1,2}$, Ivo Merks$^2$, Tao Zhang$^2$, Simon Doclo$^1$
\thanks{
This work was supported in part by a joint Lower Saxony-Israeli Project financially supported by the State of Lower Saxony, Germany, by the Cluster of Excellence 1077 Hearing4all, funded by the German Research Foundation (DFG), and a research gift from Starkey Hearing Technologies.
}}%
\address{$^1$University of Oldenburg, Department of Medical Physics and Acoustics and Cluster of Excellence\\ Hearing4All, Oldenburg, Germany\\
$^2$Starkey Hearing Technologies, Eden Prairie, MN, 55344, USA\\
}
\begin{document}
\newcommand{\beq}{\begin{equation}}
\newcommand{\eeq}{\end{equation}}

\newcommand{\vect}[1]{\mathbf{#1}}
\newcommand{\argmin}{\operatornamewithlimits{argmin}}

\newcommand{\R}{\vect{R}}
\newcommand{\Rt}{\widetilde{\vect{R}}}
\newcommand{\invRt}{\widetilde{\vect{R}}^{-1}}
\newcommand{\Rx}{\vect{R}_{\rm x}}
\newcommand{\Rz}{\vect{R}_{\rm z}}
\newcommand{\Ri}{\vect{R}_{\rm u}}
\newcommand{\Ru}{\vect{R}_{\rm u}}
\newcommand{\Run}{\vect{R}_{{\rm u},p}}
\newcommand{\Rv}{\vect{R}_{\rm v}}
\newcommand{\invRx}{\vect{R}_{\rm x}^{-1}}
\newcommand{\invRi}{\vect{R}_{\rm i}^{-1}}
\newcommand{\invRv}{\vect{R}_{\rm v}^{-1}}
\newcommand{\Ps}{{\Phi}_{\rm x}}
\newcommand{\Pint}{{\Phi}_{\rm u}}
\newcommand{\Pintn}{{\Phi}_{{\rm u},p}}
\newcommand{\Pn}{\Phi_{\rm n}}

\newcommand{\Pxlr}{\Phi_{\rm{x,LR}}}
\newcommand{\Pxl}{\Phi_{\rm{x,L}}}
\newcommand{\Pxr}{\Phi_{\rm{x,R}}}
\newcommand{\Px}{\Phi_{\rm x}}

\newcommand{\Pylr}{\Phi_{\rm{y,LR}}}
\newcommand{\Pyl}{\Phi_{\rm{y,L}}}
\newcommand{\Pyr}{\Phi_{\rm{y,R}}}
\newcommand{\Py}{\Phi_{\rm y}}

\newcommand{\Pzlr}{\Phi_{\rm{z,LR}}}
\newcommand{\Pzl}{\Phi_{\rm{z,L}}}
\newcommand{\Pzr}{\Phi_{\rm{z,R}}}
\newcommand{\Pz}{\Phi_{\rm z}}

\newcommand{\Pznlr}{\Phi_{\rm{zn,LR}}}
\newcommand{\Pznl}{\Phi_{\rm{zn,L}}}
\newcommand{\Pznr}{\Phi_{\rm{zn,R}}}
\newcommand{\Pzn}{\Phi_{\rm zn}}

\newcommand{\Pulr}{\Phi_{\rm{u,LR}}}
\newcommand{\Pul}{\Phi_{\rm{u,L}}}
\newcommand{\Pur}{\Phi_{\rm{u,R}}}
\newcommand{\Pvlr}{\Phi_{\rm{v,LR}}}
\newcommand{\Pvl}{\Phi_{\rm{v,L}}}
\newcommand{\Pvr}{\Phi_{\rm{v,R}}}
\newcommand{\Pv}{\Phi_{\rm v}}
\newcommand{\Pnlr}{\Phi_{\rm{n,LR}}}
\newcommand{\Pnl}{\Phi_{\rm{n,L}}}
\newcommand{\Pnr}{\Phi_{\rm{n,R}}}

\newcommand{\aah}{\vect{a}\vect{a}^H}
\newcommand{\abh}{\vect{a}\vect{b}^H}
\newcommand{\bah}{\vect{b}\vect{a}^H}
\newcommand{\bbh}{\vect{b}\vect{b}^H}

\newcommand{\vv}{\vect{v}}
\newcommand{\vn}{\vect{n}}
\newcommand{\vx}{\vect{x}}
\newcommand{\vu}{\vect{u}}
\newcommand{\vun}{\vect{u}_p}
\newcommand{\vy}{\vect{y}}
\newcommand{\vvh}{\vect{v}^H}
\newcommand{\vnh}{\vect{n}^H}
\newcommand{\vxh}{\vect{x}^H}
\newcommand{\vuh}{\vect{u}^H}
\newcommand{\vyh}{\vect{y}^H}

\newcommand{\vdl}{\vect{\bf D}_\mathrm{L}(\theta_i)}
\newcommand{\vdr}{\vect{\bf D}_\mathrm{R}(\theta_i)}
\newcommand{\vd}{\vect{\bf D}(\theta_i)}
\newcommand{\vdlh}{\vect{\bf D}_\mathrm{L}^H(\theta_i)}
\newcommand{\vdrh}{\vect{\bf D}_\mathrm{R}^H(\theta_i)}
\newcommand{\vdh}{\vect{\bf D}^H(\theta_i)}
\newcommand{\dl} {D_\mathrm{L}(\theta_i)}
\newcommand{\dr} {D_\mathrm{R}(\theta_i)}
\newcommand{\dRtf}{H(\theta_i)}

\newcommand{\vdls}{\vect{\bf D}_\mathrm{L}(\theta_s)}
\newcommand{\vdrs}{\vect{\bf D}_\mathrm{R}(\theta_s)}
\newcommand{\vds}{\vect{\bf D}(\theta_s)}
\newcommand{\vdlhs}{\vect{\bf D}_\mathrm{L}^H(\theta_s)}
\newcommand{\vdrhs}{\vect{\bf D}_\mathrm{R}^H(\theta_s)}
\newcommand{\vdhs}{\vect{\bf D}^H(\theta_s)}
\newcommand{\dls} {D_\mathrm{L}(\theta_s)}
\newcommand{\drs} {D_\mathrm{R}(\theta_s)}

\newcommand{\invGam}{{\mathbf\Gamma}^{-1}}
\newcommand{\Gam}{{\mathbf\Gamma}}

\newcommand{\va}{\vect{a}}
\newcommand{\vb}{\vect{b}}
\newcommand{\vah}{\vect{a}^H}
\newcommand{\vbh}{\vect{b}^H}
\newcommand{\vhl}{\vect{a}_\mathrm{L}}
\newcommand{\vhr}{\vect{a}_\mathrm{R}}
\newcommand{\vhhl}{\vect{a}_\mathrm{L}^H}
\newcommand{\vhhr}{\vect{a}_\mathrm{R}^H}
\newcommand{\vbhl}{\vect{B}_\mathrm{L}}
\newcommand{\vbhr}{\vect{B}_\mathrm{R}}
\newcommand{\vbhhl}{\vect{b}_\mathrm{L}^H}
\newcommand{\vbhhr}{\vect{b}_\mathrm{R}^H}

\newcommand{\ral}{\overline{\va}_\mathrm{L}}
\newcommand{\rar}{\overline{\va}_\mathrm{R}}
\newcommand{\rbl}{\overline{\vb}_\mathrm{L}}
\newcommand{\rbr}{\overline{\vb}_\mathrm{R}}

\newcommand{\ralh}{\overline{\va}_\mathrm{L}^H}
\newcommand{\rarh}{\overline{\va}_\mathrm{R}^H}
\newcommand{\rblh}{\overline{\vb}_\mathrm{L}^H}
\newcommand{\rbrh}{\overline{\vb}_\mathrm{R}^H}

\newcommand{\MSCvOut}{{\rm{MSC}}_{\rm v}^{\rm out}}
\newcommand{\MSCvIn}{\rm{MSC}_{\rm v}^{\rm in}}
\newcommand{\MSCvDes}{\rm{MSC}_{\rm v}^{\rm des}}
\newcommand{\SNRlIn}{\rm{SNR}_\mathrm{L}^{\rm in}}
\newcommand{\SNRrIn}{\rm{SNR}_\mathrm{R}^{\rm in}}

\newcommand{\SIRl}{\textrm{SIR}_\mathrm{L}}
\newcommand{\SIRr}{\textrm{SIR}_\mathrm{R}}
\newcommand{\SNRl}{\textrm{SNR}_\mathrm{L}}
\newcommand{\SNRr}{\textrm{SNR}_\mathrm{R}}
\newcommand{\SIRil}{\textrm{SIRin}_\mathrm{L}}
\newcommand{\SIRir}{\textrm{SIRin}_\mathrm{R}}
\newcommand{\SNRil}{\textrm{SNRin}_\mathrm{L}}
\newcommand{\SNRir}{\textrm{SNRin}_\mathrm{R}}
\newcommand{\SIRol}{\textrm{SIRout}_\mathrm{L}}
\newcommand{\SIRor}{\textrm{SIRout}_\mathrm{R}}
\newcommand{\SNRol}{\textrm{SNRout}_\mathrm{L}}
\newcommand{\SNRor}{\textrm{SNRout}_\mathrm{R}}

\newcommand{\el}{\vect{e}_\mathrm{L}}
\newcommand{\elh}{\vect{e}_\mathrm{L}^T}
\newcommand{\er}{\vect{e}_\mathrm{R}}
\newcommand{\erh}{\vect{e}_\mathrm{R}^T}
\newcommand{\rhotl}{\tilde{\rho}_\mathrm{L}}
\newcommand{\rhotr}{\tilde{\rho}_\mathrm{R}}

\newcommand{\gh}{\vect{g}^H}
\newcommand{\w}{\vect{w}}
\newcommand{\wh}{\vect{w}^H}
\newcommand{\SIR}{\textrm{SIR}}
\newcommand{\SNR}{\textrm{SNR}}
\newcommand{\wl}{\vect{w}_\mathrm{L}}
\newcommand{\wrr}{\vect{w}_\mathrm{R}}
\newcommand{\wlh}{\vect{w}_\mathrm{L}^H}
\newcommand{\wrrh}{\vect{w}_\mathrm{R}^H}
\newcommand{\C}{\vect{C}}
\newcommand{\Ch}{\vect{C}^H}
\newcommand{\bcon}{\vect{g}}
\newcommand{\bconh}{\vect{g}^H}

\newcommand{\bl}{B_\mathrm{L}}
\newcommand{\br}{B_\mathrm{R}}
\newcommand{\al}{A_\mathrm{L}}
\newcommand{\ar}{A_\mathrm{R}}
\newcommand{\rx}{\vect{r}_{\rm x}}
\newcommand{\rxh}{\vect{r}_{\rm x}^H}
\newcommand{\rxl}{\vect{r}_{\rm{x,L}}}
\newcommand{\rxlh}{\vect{r}_{\rm{x,L}}^H}
\newcommand{\rxr}{\vect{r}_{\rm{x,R}}}
\newcommand{\rxrh}{\vect{r}_{\rm{x,R}}^H}

\newcommand{\rv}{\vect{r}_{\rm v}}
\newcommand{\rvh}{\vect{r}_{\rm v}^H}
\newcommand{\rvl}{\vect{r}_{\rm{v,L}}}
\newcommand{\rvlh}{\vect{r}_{\rm{v,L}}^H}
\newcommand{\rvr}{\vect{r}_{\rm{v,R}}}
\newcommand{\rvrh}{\vect{r}_{\rm{v,R}}^H}

\newcommand{\Rn}{\vect{R}_{\rm n}}
\newcommand{\Rnl}{\vect{R}_{\rm{n,L}}}
\newcommand{\Rnr}{\vect{R}_{\rm{n,R}}}
\newcommand{\Rnlr}{\vect{R}_{\rm{n,LR}}}

\newcommand{\Ry}{\vect{R}_{\rm y}}
\newcommand{\Ryl}{\vect{R}_{\rm{y,L}}}
\newcommand{\Ryr}{\vect{R}_{\rm{y,R}}}
\newcommand{\Rylr}{\vect{R}_{\rm{y,LR}}}

\newcommand{\Rxn}{\vect{R}_{\rm xn}}
\newcommand{\Runn}{\vect{R}_{{\rm un},p}}

\newcommand{\invRy}{\vect{R}_{\rm y}^{-1}}
\newcommand{\Ryt}{\widetilde{\vect{R}}_{\rm y}}
\newcommand{\invRyt}{\widetilde{\vect{R}}_{\rm y}^{-1}}
\newcommand{\invR}{\vect{R}^{-1}}
\newcommand{\Rvt}{\widetilde{\vect{R}}_{\rm v}}
\newcommand{\invRvt}{\widetilde{\vect{R}}_{\rm v}^{-1}}

\newcommand{\sigab}{\sigma_{ab}}
\newcommand{\siga}{\sigma_{a}}
\newcommand{\sigb}{\sigma_{b}}

\ninept
\maketitle
\begin{abstract}
%
Binaural beamforming algorithms for head-mounted assistive listening devices are crucial to improve speech quality and speech intelligibility in noisy environments, while maintaining the spatial impression of the acoustic scene.
While the well-known BMVDR beamformer is able to preserve the binaural cues of one desired source, the BLCMV beamformer uses additional constraints to also preserve the binaural cues of interfering sources.
In this paper, we provide theoretical and practical insights on how to optimally set the interference scaling parameters in the BLCMV beamformer for an arbitrary number of interfering sources.
In addition, since in practice only a limited temporal observation interval is available to estimate all required beamformer quantities, we provide an experimental evaluation in a complex acoustic scenario  using measured impulse responses from hearing aids in a cafeteria for different observation intervals.
The results show that even rather short observation intervals are sufficient to achieve a decent noise reduction performance and that a proposed threshold on the optimal interference scaling parameters leads to smaller binaural cue errors in practice.
\end{abstract}
\section{Introduction}\label{sec:intro}
%
For head-mounted assistive listening devices (e.g., hearing aids, cochlear implants), algorithms that use the microphone signals from both the left and the right hearing device are effective techniques to improve speech intelligibility, as the spatial information captured by all microphones can be exploited \cite{Doclo2015,Doclo2018}. Besides reducing undesired sources and limiting speech distortion, another important objective of binaural speech enhancement algorithms is the preservation of the listener's perception of the acoustical scene, in order to exploit the binaural hearing advantage \cite{Bronkhorst:1988} and to reduce confusions due to a mismatch between acoustical and visual information.\\
To achieve binaural noise reduction with binaural cue preservation, two main concepts have been developed.
In the first concept, a common real-valued spectro-temporal gain is applied to the reference microphone signals in the left and the right hearing device \cite{Lotter2006,Grimm:2009,Kamkar2009,Kamkar2011,Reindl2013,Baumgaertel2015,Marquardt2017:2}, ensuring perfect preservation of the instantaneous binaural cues but inevitably introducing speech distortion.
The second concept, which is considered in this paper, is to apply a complex-valued filter to all available microphone signals on the left and the right hearing device using binaural extensions of spatial filtering techniques \cite{Aichner2007,Cornelis2010,Marquardt2015:3,Hadad2015:2,Hadad2016,Hadad2017,Koutrouvelis2017,Pu2017,Marquardt2018}.\\
While the well-known binaural minimum variance distortionless response (BMVDR) beamformer \cite{Doclo2015} preserves the binaural cues (i.e., the interaural level difference (ILD) and interaural time difference (ITD)) of one desired source, the binaural linearly constrained minimum variance (BLCMV) beamformer \cite{Hadad2016} is also able to preserve the binaural cues of interfering sources.
This is achievable by imposing interference scaling constraints for these sources.
It should be noted that the BMVDR and BLCMV beamformers require an estimate of the correlation matrix that should be minimized and an estimate of the \textit{relative transfer functions} (RTFs) of the desired and interfering sources.
The performance of these beamformers may significantly deteriorate in case of estimation errors. Such estimation errors occur if only short temporal observation intervals for estimation can be used, e.g., due to dynamic spatial scenarios such as moving sources or head movement.\\
In this paper, we first derive optimal values for the interference scaling parameters in the BLCMV beamformer based on the BMVDR beamformer with RTF preservation (BMVDR-RTF) \cite{Marquardt2015:3,Hadad2015:2} for an arbitrary number of interfering sources.
Secondly, since these values are optimal in the sense of noise reduction but not robust against RTF estimation errors in practice, we propose to apply an upper and lower threshold on them.
We evaluate the performance of the BMVDR beamformer and the BLCMV beamformer using the two different interference scaling parameters and measured impulse responses from hearing aids in a cafeteria \cite{Kayser2009} for several temporal observation intervals.
The results show that even rather short temporal temporal observation intervals lead to sufficient noise reduction performance and that the imposed threshold on the optimal interference scaling parameters can significantly reduce binaural cue errors.
%
\begin{figure}[t!]
\centering\includegraphics[width=6.0cm]{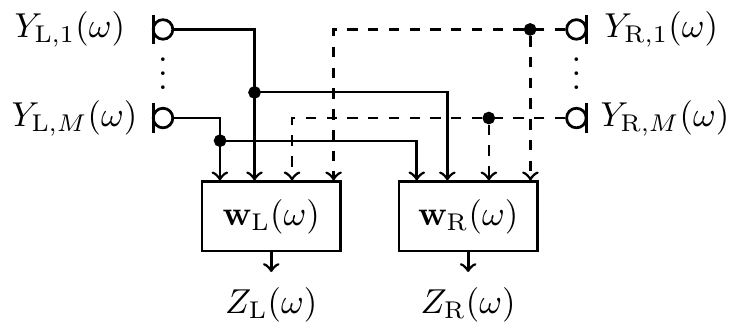}
    \caption{Binaural hearing device configuration.}
    \label{fig:config}
    \vspace{3mm}
\end{figure}
\section{Configuration and Notation}\label{sec:config}
%
Consider the binaural hearing device configuration in Fig. \ref{fig:config}, consisting of a microphone 
array with $M$ microphones on the left and the right hearing device. For an acoustic scenario with one desired source, $P$ interfering sources and incoherent background noise, the $m$-th microphone signal of the left hearing device $Y_{{\rm L},m}\left(\omega\right)$ can be written in the frequency-domain as
%
%
\begin{nospaceflalign}\label{eq:Y0}
Y_{{\rm L},m}\left(\omega\right) = X_{{\rm L},m}\left(\omega\right) + \sum_{p=1}^{P}\hspace{1mm}U_{{\rm L},p,m}\left(\omega\right) + N_{{\rm L},m}\left(\omega\right),
%
\end{nospaceflalign}
with $X_{{\rm L},m}\left(\omega\right)$ the desired speech component, $U_{{\rm L},p,m}\left(\omega\right)$ the $p$-th interference component and $N_{{\rm L},m}\left(\omega\right)$ the background noise component (e.g., diffuse noise) in the $m$-th microphone signal. The $m$-th microphone signal of the right hearing device $Y_{{\rm R},m}\left(\omega\right)$ is defined similarly.       
For conciseness we will omit the frequency variable $\omega$ in the remainder of the paper.
We define the $2M$-dimensional stacked signal vector $\mathbf{y}$ as
\begin{nospaceflalign}\label{eq:Y}
		\mathbf{y} = \left[Y_{{\rm L},1}\ldots Y_{{\rm L},M}\,
		 Y_{{\rm R},1}\ldots Y_{{\rm R},M}\right]^{T},
\end{nospaceflalign}
where $(\cdot)^T$ denotes the transpose and which can be written as 
\begin{nospaceflalign}\label{eq:Yvec}
\vy = \vx + \vv, \quad \vv = {\sum_{p=1}^P \vu_{p}} + \vn,  
\end{nospaceflalign}
where $\vx$, $\vu_p$ and $\vn$ are defined similarly as in (\ref{eq:Y}) and $\vv$ denotes the overall undesired component, i.e., interference plus background noise components. 
For the coherent desired source $S_{\rm x}$ and the coherent interfering sources $S_{{\rm u},p}$, with $p \in \{1,\dots,P\}$, the vectors $\vx$ and $\vu_p$ can be written as 
\begin{nospaceflalign}
{\bf x} = S_{\rm x}\va,\qquad {\bf u}_p = S_{{\rm u},p}\vb_p, 
\end{nospaceflalign}
with $\va$ and $\vb_p$ the \textit{acoustic transfer functions} (ATFs) between all microphones and the desired and the $p$-th interfering source, respectively. 
Without loss of generality, we choose the first microphones on the left and the right hearing device as reference microphones, i.e.,
\begin{nospaceflalign}
    Y_{{\rm L}} = {\bf e}_{\rm L}^T {\bf y}, \quad Y_{{\rm R}} = {\bf e}_{\rm R}^T {\bf y},
\end{nospaceflalign}
where ${\bf e}_{\rm L}$ and ${\bf e}_{\rm R}$ are $2M$-dimensional vectors with one element equal to $1$ and the other elements equal to $0$, i.e., ${\bf e}_{\rm L}[1]=1$ and ${\bf e}_{\rm R}[M+1]=1$. 
The correlation matrices of the background noise component, the desired speech component, the $p$-th interference component and all interference components are defined as
\begin{nospaceflalign}\label{eq:corrMat}
\Rn &= \mathop{\mathcal E}\left\{\mathbf{n}\mathbf{n}^{H}\right\},\quad
\Rx = \mathop{\mathcal E}\left\{\vx\vxh\right\} = \Ps\va\vah,\\ 
\Run &= \mathop{\mathcal E}\left\{\vun\vun^{H}\right\} = \Pintn\vb_p\vbh_p, \quad \Ru = \sum_{p=1}^{P}\Run,
\end{nospaceflalign}
where $\mathop{\mathcal E}\left\{\cdot\right\}$ denotes the expectation operator, $(\cdot)^H$ denotes the conjugate transpose and $\Ps$ and $\Pintn$ denote the \textit{power spectral density} (PSD) of the desired source and the $p$-th interfering source, respectively.
Assuming statistical independence between the components in (\ref{eq:Y0}), the correlation matrix of the microphone signals $\Ry$ can be written as  
%
\begin{nospaceflalign}\label{eq:Ry}
\hspace{-0.25cm}
\Ry = \Rx + \Ru + \Rn = \Rx + \Rv,
\end{nospaceflalign}
with $\Rv$ the correlation matrix of the overall undesired component.\\  
The output signal at the left hearing device $Z_{\rm L}$ is obtained by filtering the microphone signals with the $2M$-dimensional filter $\wl$, i.e.,
\begin{nospaceflalign}\label{eq:Z}
    Z_{\rm L} = \wlh\vy = \wlh\vx + \sum_{p=1}^{P}\wlh\vun + \wlh\mathbf{n}, 
\end{nospaceflalign}
The output signal at the right hearing aid $Z_{\rm R}$ is similarly defined.
Furthermore, we define the $4M$-dimensional filter vector ${\bf w}$ as
\begin{nospaceflalign}\label{eq:stackedweight}
{\bf w} = \lba {\bf w}_{\rm L} \\ {\bf w}_{\rm R} \ear \; .
%
\end{nospaceflalign}
The RTF vectors of the desired and the interfering sources are defined by relating the ATF vectors to the ATF of the reference microphone on the left and the right hearing device, i.e., 
\begin{nospaceflalign}\label{eq:RtfVec}
\vhl = \frac{\va}{\al},\quad\vhr = \frac{\va}{\ar},\quad
\vect{b}_{{\rm L},p} = \frac{\vb_p}{B_{{\rm L},p}},\quad\vect{b}_{{\rm R},p} = \frac{\vb_p}{B_{{\rm R},p}}.
\end{nospaceflalign}
The $2M\times P$-dimensional matrices $\vbhl$ and $\vbhr$ containing the RTF vectors of all interfering sources are defined as 
\begin{nospaceflalign}\label{eq:Bvev}
\vbhl = \lba \vect{b}_{{\rm L},1},\ldots, \vect{b}_{{\rm L},P}\ear,\quad
\vbhr = \lba \vect{b}_{{\rm R},1},\ldots, \vect{b}_{{\rm R},P}\ear.
\end{nospaceflalign}
The binaural input and output {\textit{signal-to-noise ratio}} (SNR) is defined as the ratio of the average input and output PSDs of the desired speech component and the background noise component, i.e.,
\begin{nospaceflalign}\label{eq:sinr}
{\rm SNR^{i}} = \frac{\elh\Rx\el + \erh\Rx\er}{\elh\Rn\el + \erh\Rn\er}, \quad {\rm SNR^{o}} = \frac{\wlh\Rx\wl + \wrrh\Rx\wrr}{\wlh\Rn\wl + \wrrh\Rn\wrr}.
\end{nospaceflalign}
The binaural input and output {\it signal-to-interference ratio} (SIR) is defined as the ratio of the average input and output PSDs of the desired speech component and the interference components, i.e.,
\begin{nospaceflalign}\label{eq:sinr}
{\rm{SIR}}^{\rm i} = \frac{\elh\Rx\el + \erh\Rx\er}{\elh\Ru\el + \erh\Ru\er}, \quad {\rm{SIR}}^{\rm o} = \frac{\wlh\Rx\wl + \wrrh\Rx\wrr}{\wlh\Ru\wl + \wrrh\Ru\wrr}.
\end{nospaceflalign}
The binaural input and output {\it signal-to-interference-plus-noise ratio} (SINR) is defined as the ratio of the average input and output PSDs of the desired speech component and the overall undesired component, i.e.,
\begin{nospaceflalign}\label{eq:sinr}
{\rm SINR^{i}} = \frac{\elh\Rx\el + \erh\Rx\er}{\elh\Rv\el + \erh\Rv\er}, \quad {\rm SINR^{o}} = \frac{\wlh\Rx\wl + \wrrh\Rx\wrr}{\wlh\Rv\wl + \wrrh\Rv\wrr}.
\end{nospaceflalign}
%
%
%
%
%
\section{Binaural noise reduction algorithms}\label{sec:algo}
%
In Section 3.1 and 3.2 we briefly review the BMVDR beamformer \cite{Doclo2015,Doclo2018,Cornelis2010} and the BLCMV beamformer \cite{Hadad2016}.
Based on the optimality of the BMVDR-RTF beamformer \cite{Hadad2015:2} in optimizing the SINR (or SNR) while preserving the binaural cues of all sources, in Section 3.3 we derive optimal values for the interference scaling parameters in the BLCMV beamformer in the case of an arbitrary number of interfering sources.
Furthermore, in order to achieve a robust binaural cue preservation performance in case of estimation errors of the correlation matrices and the RTF vectors (Section 3.4), we propose to threshold these interference scaling parameters.
%
\subsection{BMVDR beamformer}\label{sec:mvdr}
%
The BMVDR beamformer aims at minimizing the output PSD in both hearing devices, while preserving the desired speech component in the reference microphone signals.
The corresponding constrained optimization problem is given by 
\begin{nospaceflalign}
\min_{{\bf w}}~\wh\Rt\w \quad \text{subject to}\quad \wh{\bf C} = {\bf g},\label{eq:mvdrCost}
\end{nospaceflalign}
with
%
\begin{nospaceflalign}\label{eq:Rvt}
\Rt = \lba \R & {\bf 0}_{2M\times 2M} \\ {\bf 0}_{2M\times 2M} & \R \ear, 
\end{nospaceflalign}
with $\R$ either equal to the correlation matrix $\Ry$ of the microphone signals, the correlation matrix $\Rv$ of the overall undesired component or the correlation matrix $\Rn$ of the background noise component.
The constraint set in \eqref{eq:mvdrCost} is given by
\begin{nospaceflalign}\label{eq:MVDRconst}
{\bf C} = \lba \vhl &{\bf 0}_{2M\times 1}\\
{\bf 0}_{2M\times 1} &\vhr\ear,\qquad
{\bf g} = \lba 1 & 1\ear,
\end{nospaceflalign}
requiring the RTF vectors of the desired source.
The solution to the optimization problem in (\ref{eq:mvdrCost}) using the constraint set in (\ref{eq:MVDRconst}) is equal to \cite{Doclo2015,Cornelis2010,Veen1988} 
\begin{nospaceflalign}\label{eq:MVDR}
{\w}_{\rm MVDR,{\rm L}} = \frac{\invR\vhl}{\vhhl\invR\vhl},\quad
{\w}_{\rm MVDR,{\rm R}} = \frac{\invR\vhr}{\vhhr\invR\vhr}.
\end{nospaceflalign}
From a theoretical point of view, in the case of perfectly estimated quantities (i.e., correlation matrices and RTF vector), using $\R = \Ry$ or $\R = \Rv$ in \eqref{eq:MVDR} is optimal in the SINR sense, whereas using $\R = \Rn$ in \eqref{eq:MVDR} is optimal in the SNR sense.
While the BMVDR beamformer preserves the binaural cues of the desired source, its major drawback is the distortion of the binaural cues of the interfering sources (and background noise), such that all sources are perceived as coming from the direction of the desired source.
In practice, it should also be realized that using $\R = \Ry$ may lead to target cancellation in the case of RTF estimation errors of the desired source \cite{Veen1988} and that $\Rv$ is not straightforward to estimate.
%
\subsection{BLCMV beamformer}\label{sec:LCMV}
%
In order to also take binaural cue preservation of the interfering sources into account as well as control the amount of interference suppression, it has been proposed in \cite{Hadad2016} to add interference scaling constraints to the BMVDR beamformer, leading to the BLCMV beamformer.
This corresponds to the constrained optimization problem in \eqref{eq:MVDRconst} with the constraint set
\begin{nospaceflalign}\label{eq:LCMVcon}
%
{\bf C}_{\rm 1}\hspace{-0.05cm} =\hspace{-0.05cm} \lba\hspace{-0.05cm} \vhl &\hspace{-0.25cm}\vbhl &\hspace{-0.15cm}{\bf 0}_{2M\times 1} &\hspace{-0.15cm}{\bf 0}_{2M\times P}\\
\hspace{-0.0cm}{\bf 0}_{2M\times 1} &\hspace{-0.15cm}{\bf 0}_{2M\times P} &\hspace{-0.15cm}\vhr &\vbhr \ear\hspace{-0.05cm},\hspace{0.3cm}
{\bf g}_{\rm 1} \hspace{-0.05cm}=\hspace{-0.05cm} \lba 1  &\hspace{-0.15cm} {\boldsymbol \delta}_{\rm L} &\hspace{-0.15cm} 1 &\hspace{-0.15cm}{\boldsymbol\delta}_{\rm R} \ear,
\end{nospaceflalign}
requiring the RTF vectors of the desired source and all interfering sources.
The $P$-dimensional vectors 
%
%
${\boldsymbol \delta}_{\rm L} = \lba \delta_{{\rm L},1}\ldots\delta_{{\rm L},P} \ear$ and 
${\boldsymbol \delta}_{\rm R} = \lba \delta_{{\rm R},1}\ldots\delta_{{\rm R},P} \ear$
contain the {\it interference scaling parameters}, which control the suppression and the binaural cue preservation of the $P$ interfering sources.
The BLCMV beamformer is given by
%
\begin{nospaceflalign}\label{eq:lcmvW}
{\bf w}_{\rm LCMV} &= \invRt\C_{1}\left(\Ch_{1}\invRt\C_{1}\right)^{-1}\bconh_{1}.
\end{nospaceflalign}
Setting $\delta_{{\rm L},p} = \delta_{{\rm R},p}$ ensures binaural cue preservation of the $p$-th interfering source, while the absolute values of $\delta_{{\rm L},p}$ and $\delta_{{\rm R},p}$ directly determine the SIR improvement for the $p$-th interfering source.
From a theoretical point of view, in the case of perfectly estimated quantities (i.e., correlation matrices and RTF vectors), setting $\delta_{{\rm L},p} = \delta_{{\rm R},p} = 0$ in the BLCMV beamformer is optimal in the SIR sense, but not necessarily in the SINR or SNR sense.
Moreover, in contrast to the BMVDR beamformer, the choice of the correlation matrix $\R$ has no impact on the SINR, SNR and SIR improvement and the binaural cue preservation as these are completely determined by the interference scaling parameters.
In practice, in the case of estimation errors the choice of the correlation matrix $\R$ will obviously have an influence on the performance of the BLCMV beamformer (cf. Section 4).
%
\subsection{Interference scaling parameters}\label{sec:DeltaCon}
%
As an extension of the method presented in \cite{Marquardt2014:2} for an arbitrary number of interfering sources, in this section we propose a method to determine the interference scaling parameters that maximize the SINR or the SNR while preserving the binaural cues of the interfering sources.
To this end, we will use the BMVDR beamformer with RTF preservation \cite{Hadad2015:2}, denoted as BMVDR-RTF beamformer, which is a special case of the BLCMV beamformer.
In the BMVDR-RTF beamformer the constraints related to the interfering sources only control the binaural cue preservation while the amount of desired interference suppression is not specified, i.e.,
\begin{nospaceflalign}
\frac{\wlh\vect{b}_p}{\wrrh\vect{b}_p} = \frac{B_{{\rm L},p}}{B_{{\rm R},p}} \Rightarrow \frac{\wlh\vect{b}_{{\rm L},p}}{\wrrh\vect{b}_{{\rm R},p}} = 1,
\end{nospaceflalign}
leading to the constraint set  
\begin{nospaceflalign}\label{eq:ConRTF}
\hspace{-0.15cm}
{\bf C}_2 &= \lba \vhl &\vbhl &{\bf 0}_{2M\times 1}\\
{\bf 0}_{2M\times 1} &-\vbhr &\vhr \ear,\quad
{\bf g}_2 = \lba 1  & {\bf 0}_{1\times P} & 1\ear.
\end{nospaceflalign}
The BMVDR-RTF beamformer is given by \cite{Hadad2015:2}
%
\begin{nospaceflalign}\label{eq:Wrtf}
{\bf w}_{\rm RTF} &= \invRt\C_2\left(\Ch_2\invRt\C_2\right)^{-1}\bconh_2,
\end{nospaceflalign}
and either maximizes the SINR ($\R=\Ry$ or $\R=\Rv$) or the SNR ($\R=\Rn$), while preserving the binaural cues of all sources.\\
Hence, the optimal interference scaling parameters for the BLCMV beamformer (in the SINR or SNR sense) can be determined as
\begin{equation}
\boxed{
  \delta^{\rm opt}_{p} = \delta_{{\rm L},p} = \delta_{{\rm R},p} = {\bf w}_{\rm RTF,L}^H\vb_{{\rm L},p} = {\bf w}_{\rm RTF,R}^H\vb_{{\rm R},p}
  }
\end{equation}
However, using the optimal interference scaling parameters may lead to problems in practice due to estimation errors of the correlation matrices and RTF vectors.
More in particular, in the case of SINR maximization, the corresponding interference scaling parameters may be rather small, leading to a decreased binaural cue preservation performance (cf. simulations in Section 4).
On the other hand, in the case of SNR maximization, the corresponding interference scaling parameters may be rather large, depending on the position of the interfering source, leading to an unsatisfying SINR improvement.
Hence, we propose to enforce an upper and lower threshold on the optimal interference scaling parameters, i.e.,
\begin{nospaceflalign}
\label{eq:delta_thr}
\delta_p^{\rm thr} = 
\begin{cases}
|\delta_p^{\rm opt}|,\quad\,\, \textrm{if} &\delta^{\rm min} < |\delta_p^{\rm opt}| < \delta^{\rm max},\\
\delta^{\rm min},\quad\, \textrm{if} &|\delta_p^{\rm opt}| \leq \delta^{\rm min},\\
\delta^{\rm max},\quad \textrm{if} &|\delta_p^{\rm opt}| \geq \delta^{\rm max}.\\
\end{cases}
\end{nospaceflalign}
The thresholds have been experimentally obtained as $\delta^{\rm min} = 0.2$ and $\delta^{\rm max} = 0.4$, limiting the theoretically possible SIR improvement for each interfering source between \SI{8}{\decibel} and \SI{14}{\decibel}.    
%
%
\subsection{Estimation of correlation matrices and RTFs}\label{sec:RtfDoa}
%
All considered binaural beamformers require an estimate of the RTF vectors $\va_{\rm L}$ and $\va_{\rm R}$ of the desired source (cf. \eqref{eq:RtfVec}).
In addition, the BLCMV and BMVDR-RTF beamformers require an estimate of the RTF vectors $\vb_{{\rm L},p}$ and $\vb_{{\rm R},p}$ of each interfering source.
In this paper, we will estimate these RTFs using the covariance whitening approach \cite{Markovich2009,Markovich2015}, which is based on the generalized eigenvalue decomposition (GEVD) of the speech + noise correlation matrix $\Rxn=\Rx + \Rn$ and the background noise correlation matrix $\Rn$ or the GEVD of the interference + noise correlation matrix $\R_{\mathrm{v},p} = \Run + \Rn$ and $\Rn$.
While $\Rn$ can be estimated exploiting the assumed stationarity of the background noise, estimating $\Rxn$ and $\R_{\mathrm{v},p}$ from the available mixture is not straightforward.
Due to limited source activity and possible spatial changes of the acoustic scenarios, the \textit{temporal observation interval} that is available in practice for estimating these correlation matrices is typically limited.
We assume that the correlation matrix $\Rxn$ can be estimated from an observation interval consisting of $T_L$ frames (corresponding to $L$ seconds) where only the desired source and the background noise are active, i.e.,
\begin{equation}
	\hat{\mathbf{R}}_{\rm xn} = \frac{1}{T_L}\sum_{t=1}^{T_L} \bigg(\mathbf{x}(t) + \mathbf{n}(t)\bigg)\bigg(\mathbf{x}(t) + \mathbf{n}(t)\bigg)^H,
\end{equation}
where $t$ is the frame index. Similarly, we assume that the correlation matrix $\mathbf{R}_{{\rm v},p}$ can be estimated from an observation interval of $T_L$ frames where only the $p$-th interfering source and the background noise are active.
\section{Experimental Results}\label{sec:eval}
\begin{figure}[t!]
	\includegraphics[width=\linewidth]{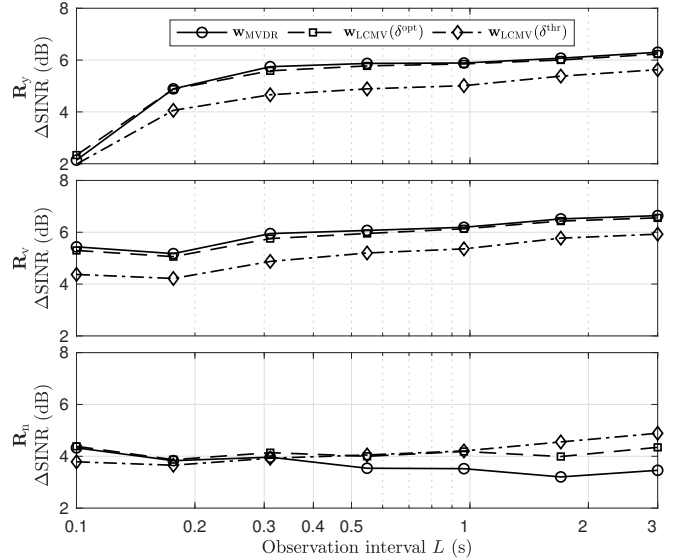}
	\vspace{-4mm}
	\caption{SINR improvement for different temporal observation intervals for $\R = \Ry$ (top), $\R = \Rv$ (mid) and $\R = \Rn$ (bottom).}
	\label{fig:results}
\end{figure}

\begin{figure}[t!]
	\centering
	\includegraphics[width=\linewidth]{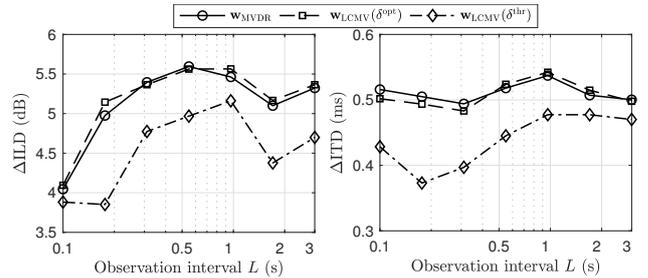}
	   \vspace{-5mm}
	\caption{Binaural cue errors of the first interfering source ($\R=\Rv$).}
	\label{fig:cues}
\end{figure}

%
In this section, we experimentally investigate the effect of the temporal observation interval on the performance of the BMVDR beamformer ($\mathbf{w}_{\rm MVDR}$) and the BLCMV beamformer using either the optimal interference scaling parameters ($\mathbf{w}_{\rm LCMV}(\delta^{\rm opt})$) or the proposed thresholded interference scaling parameters ($\mathbf{w}_{\rm LCMV}(\delta^{\rm thr})$) (cf. Section \ref{sec:DeltaCon}).\\
We consider three different acoustic scenarios comprising of one desired source, one or two interfering sources and diffuse background noise (cf. Table \ref{tab:Ex} for source positions).
The desired source was a male German speaker, the first interfering source was a male Dutch speaker and the second interfering source was a male English speaker.
The desired speech and interference components were generated by convolving the desired and interfering source signals with measured impulse responses of binaural behind-the-ear hearing aids mounted on a dummy head in a cafeteria ($T_{60} \approx \SI{1250}{ms}$)\cite{Kayser2009}, with $M=2$ microphones per hearing aid.
For background noise we used real ambient noise recorded in the same
cafeteria with the same setup. The sampling frequency was \SI{16}{kHz}. All 
signals start with \SI{2}{s} of noise-only, followed by about \SI{20}{s} of all sources 
being active.
The broadband input SNR was set to \SI{5}{\dB} and the SIRs were set to \SI{0}{\dB}.\\
The noise correlation matrix $\Rn$ was estimated using the \SI{2}{s} noise-only segment.
To estimate the correlation matrices $\Ry$, $\Rv$, $\Rxn$ and $\R_{{\rm v},p}$, we considered different temporal observation intervals (starting at \SI{2}{s}), whose length $L$ ranged between \SI{0.1}{s} and \SI{3}{s}.
To estimate the correlation matrices $\Rv$, $\Rxn$ and $\R_{\mathrm{v},p}$ the algorithm had access to the respective mixtures.
The RTF vectors of the desired source and the interfering source(s) were then calculated based on these estimated correlation matrices (cf. Section 3.4).
Please note that shorter temporal observation intervals correspond to larger estimation errors.\\
The microphone signals were processed using a weighted overlap-add framework with a block length of 256 with 50\% overlap and a square-root Hann window.
The BMVDR and BLCMV beamformers were calculated using three different correlation matrices, i.e., $\R=\Ry$ (maximizing SINR with possible target cancellation), $\R=\Rv$ (maximizing SINR) and $\R=\Rn$ (maximizing SNR).
The filters were used as fixed filters over the whole signal.\\
As performance measures we used the binaural SINR improvement and the binaural cues errors, i.e., ILD and ITD errors, that we calculated using a model of binaural auditory processing \cite{Dietz2012}.
All performance measures were averaged over all frequencies and all acoustic scenarios.\\
%
Figure \ref{fig:results} depicts the SINR improvement for different lengths of the temporal observation interval and for different correlation matrices, while Figure \ref{fig:cues} depicts the binaural cue errors of the first interfering source for the same temporal observation intervals and $\R = \Rv$.
First, it can be observed that when using $\Ry$ or $\Rv$ the SINR improvement is generally larger than when using $\Rn$. This is expected because using the noise correlation matrix $\Rn$ is maximizing the SNR and not the SINR.
Second, when using $\Ry$ or $\Rv$, an apparent difference can be seen for small observation intervals below \SI{200}{ms}. The small observation intervals lead to larger estimation errors for the correlation matrices and hence also for the RTF vectors, such that the drop in SINR improvement observed when using $\Ry$ is probably attributed to target cancellation. For longer observation intervals and hence smaller estimation errors, the difference between using $\Ry$ and $\Rv$ is smaller.
As expected, the SINR improvement of the BLCMV beamformer using the thresholded interference scaling parameters $\delta^{\rm thr}$ is smaller than for the BLCMV beamformer using the optimal interference scaling parameters $\delta^{\rm opt}$. Although, looking at the binaural cue errors, using $\delta^{\rm thr}$ in the BLCMV beamformer leads to much better binaural cue preservation, while using $\delta^{\rm opt}$ leads to similar binaural cue errors as for the BMVDR beamformer.
This difference is especially visible for the ITD error at small observation intervals and is also confirmed by informal listening tests.
Third, when using $\Rn$, the BLCMV beamformer outperforms the BMVDR beamformer for longer observation intervals above $\SI{300}{ms}$ because of the additional constraints. Additionally, using $\delta^{\rm thr}$ in the BLCMV beamformer apparently leads to marginally better SINR improvement in this case.
Because $\Rv$ is in practice very hard to accurately estimate, it should be recommended to use $\Rn$ when short observation intervals are required (e.g., in dynamic acoustic scenarios) and to use $\delta^{\rm thr}$ in the BLCMV beamformer to prevent binaural cue errors.
%
\begin{table}[t!]
\begin{center}
  \begin{tabular}{| c | c | c | c |}
  \hline
  	Scenario &   1 &   2 &  3\\\hline
  		$\text{Desired}$  & $-35^{\circ}$ & $0^{\circ}$ & $0^{\circ}$\\\hline
  	  $\text{Interfering}$ & $150^{\circ}$ &$-35^{\circ}$ & $-35^{\circ}$, $150^{\circ}$\\\hline
  	\end{tabular}
   \caption{Spatial scenarios ($0^{\circ}$: frontal direction. $-90^{\circ}$: left hand side. $90^{\circ}$: right hand side).}
   \vspace{-5mm}
  \label{tab:Ex}
\end{center}
  \end{table}
  %
%
\vspace{-4mm}
\section{Conclusions}\label{sec:concl}
\vspace{-0.2cm}
In this paper, we proposed optimal values for the interference scaling parameters in the BLCMV beamformer for an arbitrary number of interfering sources based on the BMVDR-RTF beamformer.
We showed how to set these parameters in practice such that a robust performance in the case of estimation errors can be achieved.
Evaluation results in a complex acoustic scenario showed that even short temporal observation intervals for estimating the required correlation matrices and RTF vectors are sufficient to achieve a decent noise reduction performance and binaural cue preservation.               
\bibliographystyle{IEEEbib}
\bibliography{/Users/nico/Documents/Library/literature2}
\end{document}